\documentclass[a4paper, 11pt]{article}

\usepackage{graphicx}
\usepackage{amsmath}
\usepackage{amsfonts}
\usepackage{amssymb}
\usepackage{authblk}
\usepackage[utf8]{inputenc}
\usepackage{tikz}
\usetikzlibrary{arrows,backgrounds,snakes,patterns}
\usetikzlibrary{shapes,arrows,chains}
\usepackage{verbatim}
\usepackage{booktabs}
\usepackage{multirow}
\usepackage[flushleft]{threeparttable}

\begin{document}

\title{Multi-task Deep Neural Networks in Automated Protein Function Prediction\footnote{A short version of this manuscript was accepted for oral presentation at ISMB/ECCB 2017 Function-COSI meeting}}

\author[1,2]{Ahmet Sureyya Rifaioglu \footnote{arifaioglu@ceng.metu.edu.tr}}
\author[3,4]{Tunca Doğan \footnote{tdogan@ebi.ac.uk}}
\author[3]{Maria Jesus Martin \footnote{martin@ebi.ac.uk}}
\author[4]{Rengul Cetin-Atalay \footnote{rengul@metu.edu.tr}}
\author[1]{Mehmet Volkan Atalay \footnote{vatalay@metu.edu.tr}}

\affil[1]{Department of Computer Engineering, Middle East Technical University, Ankara, 06800, Turkey}
\affil[2]{Department of Computer Engineering, İskenderun Technical University, Hatay, 31200, Turkey}
\affil[3]{European Molecular Biology Laboratory, European Bioinformatics Institute (EMBL-EBI), Hinxton, Cambridge, CB10 1SD, UK}
\affil[4]{CanSyL, Graduate School of Informatics, Middle East Technical University, Ankara, 06800, Turkey}

\maketitle

\begin{abstract}
\textbf{Background:} In recent years, deep learning algorithms have outperformed the state-of-the art methods in several areas such as computer vision, speech recognition thanks to the efficient methods for training and for preventing overfitting, advancement in computer hardware and the availability of vast amount data. The high performance of multi-task deep neural networks in drug discovery has attracted the attention to deep learning algorithms in the bioinformatics area. Protein function prediction is a crucial research area where more accurate prediction methods are still needed.  Here, we proposed a hierarchical multi-task deep neural network architecture based on Gene Ontology (GO) terms as a solution to the protein function prediction problem and investigated various aspects of the proposed architecture by performing several experiments.\\
\textbf{Results:} First, we showed that there is a positive correlation between the performance of the system and the size of training datasets. Second, we investigated whether the level of GO terms on the GO hierarchy is related to their performance. We showed that there is no relation between the depth of GO terms on the GO hierarchy (i.e. general/specific) and their performance. In addition, we included all annotations to the training of a set of GO terms to investigate whether including noisy data to the training datasets change the performance of the system. The results showed that including less reliable annotations in training of deep neural networks increased the performance of the low performed GO terms, significantly. Finally, we evaluated the performance of the system using hierarchical evaluation method. Mathews correlation coefficients was calculated as 0.75, 0.49 and 0.63 for molecular function, biological process and cellular component categories, respectively.\\
\textbf{Conclusions:} We showed that deep learning algorithms have a great potential in protein function prediction area. We plan to further improve the DEEPred by including other types of annotations from various biological data sources. Finally, we plan to construct DEEPred as an open access online tool.

\end{abstract}

\section{Background}
\label{sec:back}
Functional annotation of proteins is a crucial research area for understanding molecular mechanism of living-beings, identification of disease-causing functional changes and for discovering novel drugs. Traditionally, functions of proteins are determined by \textit{in vivo} experiments and expert curators annotate gene and protein records in biological databases using the information in the literature produced by these experiments. However, experimental and manual curation efforts are not feasible due to the continuous growth of gene and protein sequence data [1]. Therefore, accurate computational methods have been sought to help annotating functions of proteins.
\\\\
The Gene Ontology (GO) Consortium provides a controlled vocabulary to classify attributes of proteins based on representative terms, referred as ``GO terms'' [2]. Gene Ontology system divides the protein functions into three main categories: molecular function, biological process and cellular component. The functions are represented in a directed acyclic graph (DAG) structure based on inheritance relationships among the terms. Several GO term-based protein function prediction methods have been proposed in the last decade to automatically annotate proteins using machine learning and statistical analysis techniques [3-7]. However, considering the prediction performances of the current methods, it can be stated that there is still room for significant improvements in protein function prediction area. For example, Critical Assessment of Protein Function Annotation (CAFA) is an initiative, whose aim is the large-scale evaluation of protein function prediction methods based on GO terms [8,9] and the results of the first two CAFA challenges showed that protein function prediction is still a challenging area.
\\\\
Neural networks are a set of nonlinear classifiers that are inspired from biological nervous systems which take feature vectors as input and apply nonlinear transformations. Neural networks consist of an input layer, an output layer and one or more intermediate layers called hidden layers. Each layer includes a number of nodes which are connected to the nodes of the next layer via a system of weighted connections. Deep learning (i.e. deep neural networks - DNNs) algorithms can be considered as a collection of artificial neural networks that have multiple hidden layers, which take low level input features as input and build more and more complex features at each subsequent layer. Deep learning algorithms became popular in recent years thanks to the improvements in the computational power, which made the practical applications possible. Furthermore, they became an industry standard in fields such as computer vision and speech recognition [10-14]. Recently, it was shown that deep learning algorithms outperformed the state-of-the-art methods in various research areas including bioinformatics and cheminformatics [15-19]. With the development of new techniques to avoid problems such as overfitting, application of deep neural networks became more popular and feasible. Multi-task deep neural networks is designed for classification of instances for multiple tasks in a single model [20]. In multi-task deep neural networks, after a number of iterations, the outputs of the final layer of the deep neural network is fed to a non-linear function in order to calculate the probability of the query instance to have the corresponding task. Applications of multi-task deep neural networks provided significant performance increase on ligand-based drug discovery, which is similar to the protein function prediction in terms of the problem definition [19,21]. In drug discovery, the aim is to find possible interacting drug like compounds  for a given protein target, where each protein may have more than one ligand [22]. In protein function prediction, the aim is to find possible GO term associations for a given protein where each protein may have multiple functions. Therefore, both problems can be formalized in a similar manner.  
\\\\
Several important properties of deep learning architectures have been reported, making DNNs suitable to be applied to the protein function prediction problem. First of all, deep learning algorithms inherently build relationships between multiple targets, therefore they are suitable for multi-task learning by building complex features from the raw input data in a hierarchical manner. Secondly, shared hidden units among the targets enhance the prediction results of the targets having low number of training samples, therefore, had positive impact on performance significantly. Protein function prediction can be considered as a multi-label classification since each protein may have multiple functional associations and it can be formulated as multi-task learning where a single model is created to predict multiple protein functions at once. To the best of our knowledge, as of today, deep learning algorithms have not been applied to the large-scale protein function prediction problem.
\\\\
In this study, we propose a hierarchical deep learning solution, DEEPred, in order to automatically predict the functions of proteins using GO terms. In machine leaning applications, the structure of the computational systems often require problem-specific data pre-processing and post-processing steps to provide reliable predictions [23,24]. Here, we also present a hierarchical evaluation method as a post-processing of predictions, based on the structure of GO DAG. DEEPred provides a solution to GO term-based protein function prediction problem using multi-task deep neural networks with the aim of producing more accurate functional predictions for target protein sequences.

\section{Methods}
\label{sec:methods}
\subsection {Dataset Construction}
Training dataset was created using UniProtKB/Swiss-Prot database protein sequences. UniProt supports each functional annotation with one of the 21 different evidence codes, which indicate the source of the particular annotation. In this study, we used annotations with manual curation or experimental evidences (i.e. EXP, IDA, IPI, IMP, IGI and IEP), which are considered to be highly reliable. After the extraction of the annotations with manual/experimental evidence codes, we propagated the annotations to the parents of the corresponding GO term according to the ``true path rule'', which defines an inheritance relationship between GO terms [2]. Proteins that were annotated with the corresponding GO term or with one of its children terms were included in the positive training dataset of the corresponding GO term.
\\
We constructed multiple training datasets based on the number of protein associations of GO terms. For example, one of our training dataset includes all GO terms that have more than or equal to 50 protein associations. We created six different datasets where GO terms in each dataset have more than 50, 100, 200, 300, 400 and finally 500 protein associations. For each training dataset, we applied a hierarchical training method individually, which is explained in the following section.
\subsection {DEEPred Architecture for Protein Function Prediction}
In the first step of the training phase of DEEPred, protein sequences that were associated by each GO term were determined. We used a modified version of subsequence profile map (SPMap) method to generate feature vectors for proteın sequences [25]. SPMap method consists of three main modules, which are feature extraction module, clustering module and probabilistic profile construction module. In the feature extraction module, all fixed-length subsequences are extracted from positive training sequences. In the clustering module, extracted subsequences are grouped using a clustering method similar to hierarchical clustering based on BLOSUM-62 matrix for a specified similarity threshold. Once subsequences are clustered, obtained clusters are transformed into probabilistic profiles. Finally, protein sequences are converted into feature vectors based on the distribution of their subsequences over the generated probabilistic profiles. The original SPMap method constructs a profile for each GO term individually, using positive and negative training sequences of the corresponding GO term. This results in protein feature vectors with varying sizes. In this study, we modified the SPMap algorithm and generated a single reference probabilistic profile using all training sequences that were annotated by all GO terms belonging to a GO category. Subsequently, for each GO category, all training and test sequences were converted into feature vectors based on the generated reference probabilistic profile of the corresponding category. Therefore, each protein sequence was represented by a fixed-dimensional feature vector for all models in a GO category. 

\begin{figure}[h]

  \centering
    \includegraphics[width=1.00\textwidth]{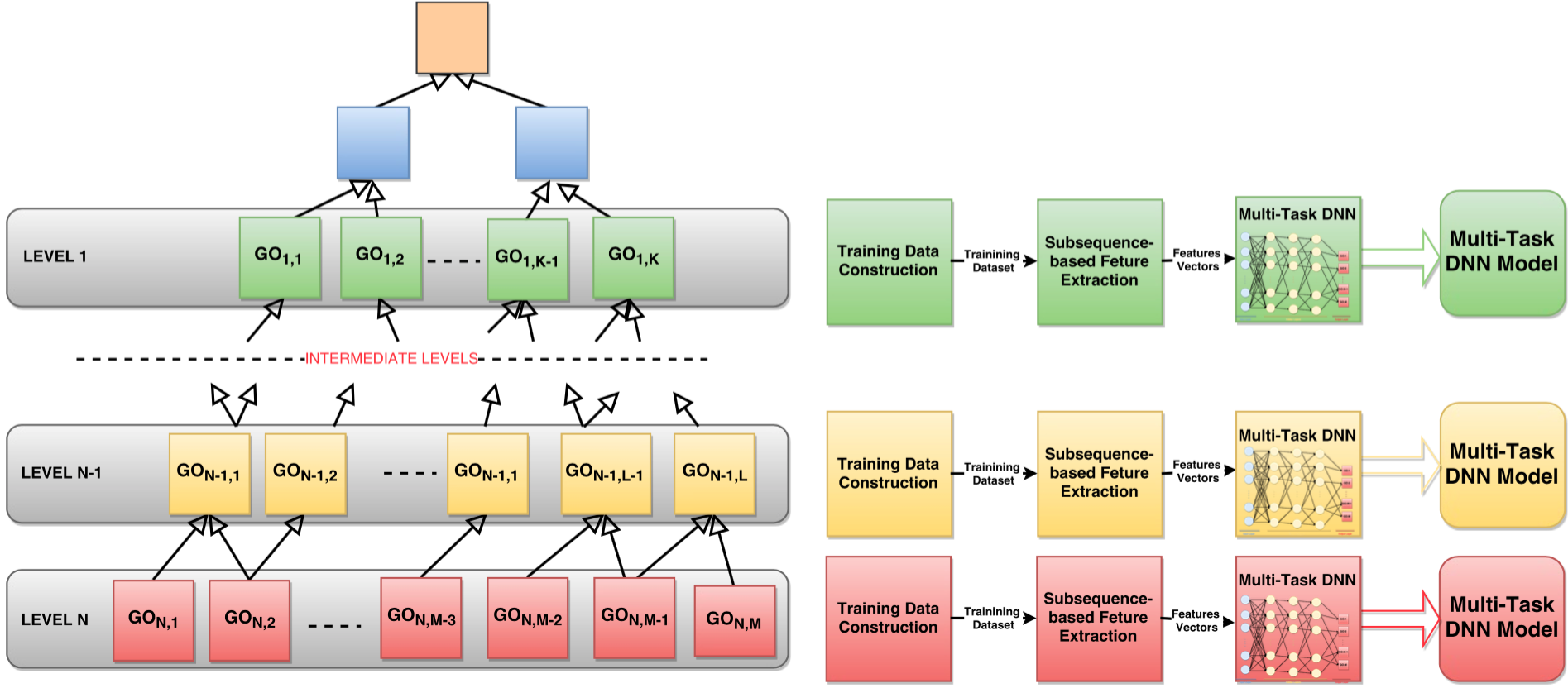}
      \caption{An illustration of DEEPred Architecture on directed acyclic graph of hypothetical GO category. We omitted general GO terms on GO directed acyclic graph.}
\end{figure}

After obtaining the feature vectors, we applied the hierarchical training procedure as follows: levels of GO terms were first identified on GO DAG for each GO category. GO terms were separated into groups based on their levels where each group corresponds to GO terms belonging to a specific level on the GO hierarchy. The main objective of this approach is to create a multi-task deep neural network model for each level (Figure 1). This way, only the functional terms with the same level of specificity (i.e. semantically comparable) are included in the same model. In some cases, number of protein associations of GO terms within a level were highly variable; therefore, we created subgroups to avoid bias, where each subgroup included the GO terms having similar number of annotations. This procedure resulted in the generation of 269 different models for all GO categories. Following the generation of the models, each model was trained using the feature vectors of the proteins annotated with the corresponding GO terms of that model.
\\
An example multi-task deep neural networks is shown in Figure 2. Here, a task corresponds to a GO term, therefore, when a query sequence is fed to our trained multitask deep neural networks models as input, we obtain probabilities for the query protein to be associated with the corresponding GO terms (i.e. to possess the function defined by the corresponding GO terms), simultaneously. We omitted some of the GO terms on the top of the GO hierarchy which are extremely generic and non-informative (e.g. GO:0005488 - Binding).
\begin{figure}[h]

  \centering
    \includegraphics[width=0.80\textwidth]{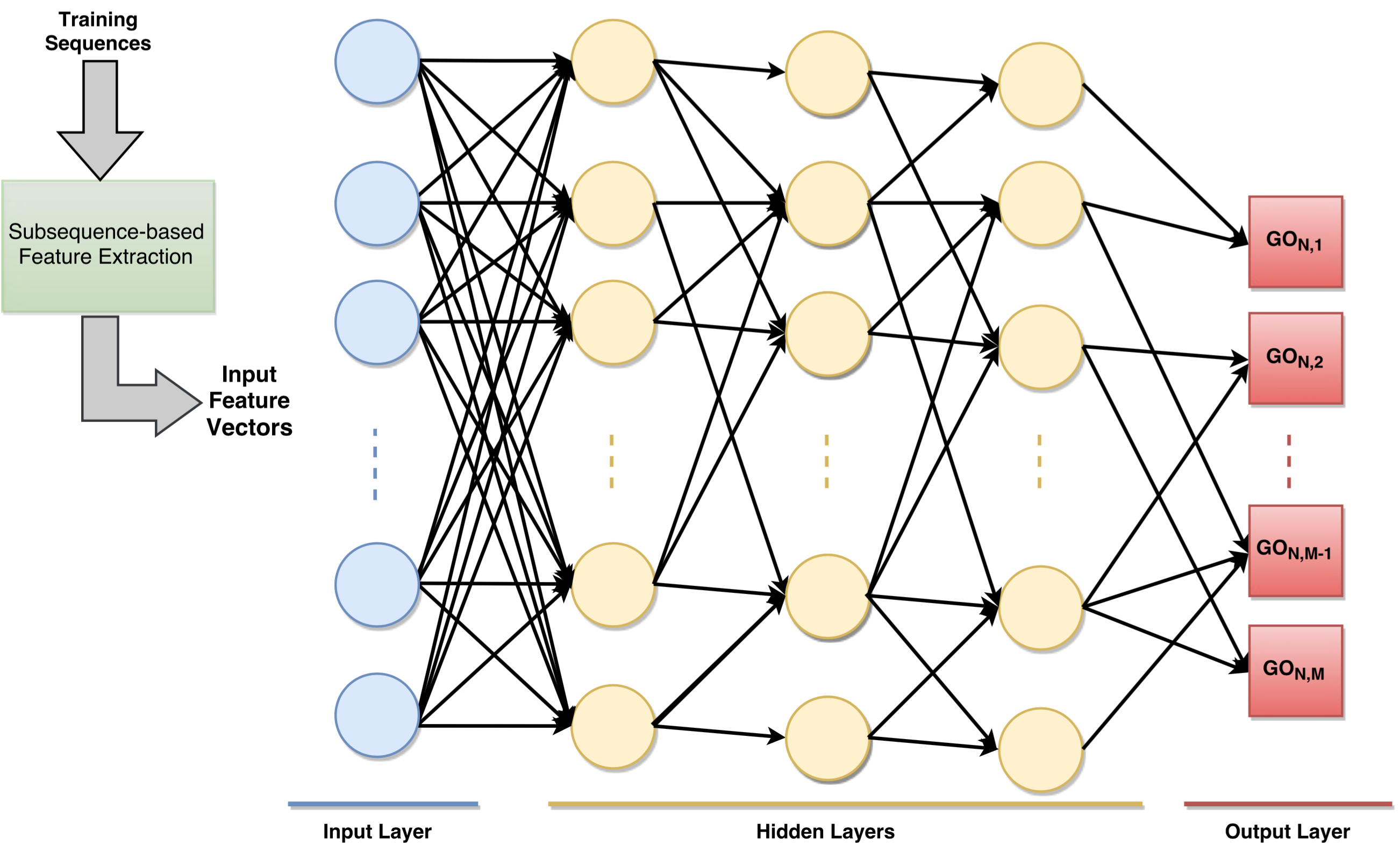}
      \caption{Multi-task deep neural network model for the Nth level on a GO category. There are M GO terms in this level and each task corresponds to a GO term on the corresponding level.  First a multi-task deep neural network model was created using features extracted from training sequences that were annotated by the GO terms to be trained. Subsequently, test sequences were fed to the system to get prediction scores for each GO term. }
\end{figure}
One of the main problems of deep learning algorithms is overfitting. Several approaches were proposed to avoid overfitting during the training of deep neural networks [26,27]. One of the most popular methods to avoid overfitting is the dropout technique [27]. Dropout technique randomly removes some neurons from different layers, along with their connections, with the aim of obtaining a more generalized model. In this study, we trained our multi-task deep neural using the dropout technique.
\\
In this study, we created several multi-task feed-forward deep neural network models with different parameters for number of hidden layers, number of neurons, learning rate and drop-out rate and we used the best performing models for each level. Number of neurons at each hidden layer ranged from 500 to 5000. Learning rate parameters ranged from 0.001 to 0.1 and drop-out rate parameters ranged from 0.3 to 0.8. Construction of multiple models allowed us to optimize the parameters for each model in order to increase the classification performance. We used TensorFlow for training models and all computations were distributed on 2500 CPU cores [28]. 
\subsection{Defining the thresholds}
DEEPred calculates a score for each trained GO term within a model, which represents the probability of an input protein possessing the function particularly defined by the corresponding GO term. Therefore, we needed to determine thresholds, to be able to indicate that the input protein received a prediction (i.e. when its score exceeds the threshold). For this purpose, we calculated F-score values for different thresholds to determine the performance of the system. We considered each GO term separately within a model and determined an individual threshold for each GO term. Subsequently, we presented average performances of GO terms for each training set.
\subsection{Hierarchical Evaluation of Overall System Performance Based on Structure of the System}
We implemented a methodology to provide predictions considering all levels of GO hierarchy, at the same time (i.e. considering the scores received for the parents of a GO term as well). This way, we aimed to reduce the potential false positive hits. For this purpose, first we topologically sorted the GO DAG for each category and determined all possible paths from each GO term to the root of the corresponding category. When a protein sequence was given as input to our method, it was first converted into a feature vector and fed to all trained models to obtain the prediction scores. Subsequently, starting from the most specific level of GO, we checked whether the prediction score of the query protein was greater than the previously calculated thresholds. If the prediction score of a target GO term is greater than its threshold, then we check the scores of its parents on all paths to the root. If the scores of the majority of parents are greater than the calculated thresholds, we present the case as a positive prediction (Figure 3). This way, we provided a GO term prediction only when the corresponding prediction is consistent with the scores of its parent terms.
\begin{figure}[h]

  \centering
    \includegraphics[width=0.90\textwidth]{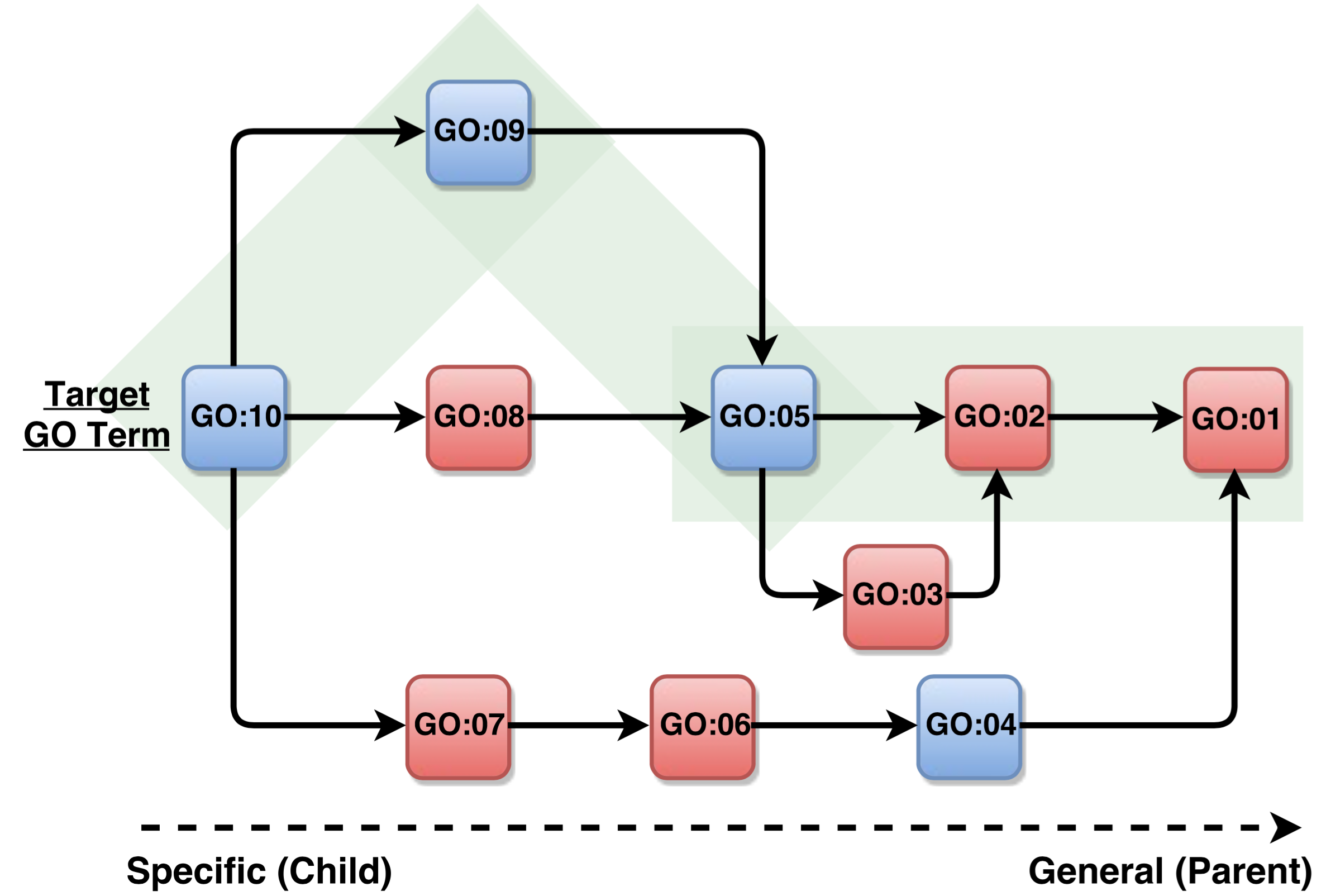}
      \caption{ Calculating the prediction for a query protein sequence for GO:1 on a hypothetical GO DAG. Each node corresponds to a GO term. When a query protein is given as input to the system, first a prediction score is calculated for each trained GO term for the query protein. The blue colored GO terms represent the GO terms whose prediction scores are over the pre-calculated threshold values whereas the red colored GO terms represent the GO terms whose prediction scores are below the pre-calculated threshold values. To provide a prediction for the target GO term, we search for all possible paths from the target GO term to the root of the GO DAG. If the prediction scores of the majority of the GO terms on one of the paths are higher than the pre-calculated thresholds (in the figure, shaded green path), we present the target GO term as a positive prediction for the input sequence.}
\end{figure}

\subsection {Evaluation Metrics}

We used F-score, precision and recall measures to evaluate performances of the system. TP, FP, TN and FN represents true positive, false positive, true negative and false negative, respectively. 

\begin{equation}
    Precision = \frac{TP}{TP+FP}
\end{equation}
\begin{equation}
    Recall = \frac{TP}{TP+FN}
\end{equation}
\begin{equation}
    Recall = \frac{{2}\times{Precision}\times{Recall}}{Precision+Recall}
\end{equation}
To evaluate overall system performance of the system using the proposed hierarchical evaluation method, we used Matthews correlation coefficient (MCC).
\begin{equation}
    MCC = \frac{ \left( {TP}\times{TN} \right) - \left( {FP}\times{FN} \right)}{ \sqrt{ {\left( {TP}+{FP} \right)} \times {\left( {TP}+{FN} \right)} \times {\left( {TN}+{FP} \right)} \times {\left( {TN}+{FN} \right)}}}
 \end{equation}
\pagebreak
\section{Results}
\label{sec:res}
\subsection{Training Dataset Construction}
We constructed 6 different training datasets based on the annotated protein counts of GO terms as described in Methods Section. The number of GO levels, the total number GO terms and the total number of annotations for training dataset (annotations with manual experimental evidence codes) and for all evidence codes are presented in Table 1. Number of levels decreases as the number of annotation count increases, since specific GO terms have less number of annotations.
\begin{table}[h]
\centering
\caption{Training dataset statistics for the sets created using only the annotations with manual experimental evidence codes and the sets created using the annotations with all evidence codes.}
\label{table-1}
\begin{tabular}{cl|r|r|r|r|r|r|}
\cline{3-8}
\multicolumn{1}{l}{} &  & \multicolumn{6}{c|}{Training Dataset Statistics} \\ \cline{3-8} 
\multicolumn{1}{l}{} &  & \multicolumn{3}{c|}{\begin{tabular}[c]{@{}c@{}}Annotations with only manual \\ experimental evidence codes\end{tabular}} & \multicolumn{3}{c|}{\begin{tabular}[c]{@{}c@{}}Annotations with \\ all evidence codes\end{tabular}} \\ \cline{2-8} 
\multicolumn{1}{l|}{} & \multicolumn{1}{c|}{\begin{tabular}[c]{@{}c@{}}Annotation\\ Count\end{tabular}} & \multicolumn{1}{c|}{\begin{tabular}[c]{@{}c@{}}\# of \\ Levels\end{tabular}} & \multicolumn{1}{c|}{\begin{tabular}[c]{@{}c@{}}\#  of\\ GO terms\end{tabular}} & \multicolumn{1}{c|}{\begin{tabular}[c]{@{}c@{}}\# of \\ Annotations\end{tabular}} & \multicolumn{1}{c|}{\begin{tabular}[c]{@{}c@{}}\# of \\ Levels\end{tabular}} & \multicolumn{1}{c|}{\begin{tabular}[c]{@{}c@{}}\#  of\\ GO terms\end{tabular}} & \multicolumn{1}{c|}{\begin{tabular}[c]{@{}c@{}}\# of \\ Annotations\end{tabular}} \\ \hline
\multicolumn{1}{|c|}{\multirow{5}{*}{MF}} & \begin{math} \geqslant \end{math} 50 & 9 & 627 & 229 400 & 14 & 2 143 & 6 372 488 \\ \cline{2-8} 
\multicolumn{1}{|c|}{} & \begin{math} \geqslant \end{math} 100 & 9 & 390 & 215 391 & 14 & 1 559 & 6 330 882 \\ \cline{2-8} 
\multicolumn{1}{|c|}{} & \begin{math} \geqslant \end{math} 200 & 8 & 225 & 192 020 & 14 & 1 127 & 6 270 019 \\ \cline{2-8} 
\multicolumn{1}{|c|}{} & \begin{math} \geqslant \end{math} 300 & 8 & 164 & 177 521 & 14 & 902 & 6 214 924 \\ \cline{2-8} 
\multicolumn{1}{|c|}{} & \begin{math} \geqslant \end{math} 400 & 7 & 137 & 168 562 & 14 & 771 & 6 169 455 \\ \cline{2-8} 
\multicolumn{1}{|c|}{} & \begin{math} \geqslant \end{math} 500 & 7 & 116 & 158 945 & 13 & 675 & 6 126 218 \\ \hline
\multicolumn{1}{|c|}{\multirow{5}{*}{BP}} & \begin{math} \geqslant \end{math} 50 & 12 & 3 030 & 1 492 883 & 16 & 6 545 & 17 056 227 \\ \cline{2-8} 
\multicolumn{1}{|c|}{} & \begin{math} \geqslant \end{math} 100 & 11 & 1 812 & 1 408 185 & 15 & 4 641 & 16 921 060 \\ \cline{2-8} 
\multicolumn{1}{|c|}{} & \begin{math} \geqslant \end{math} 200 & 10 & 1 104 & 1 307 068 & 15 & 3 224 & 16 720 787 \\ \cline{2-8} 
\multicolumn{1}{|c|}{} & \begin{math} \geqslant \end{math} 300 & 10 & 762 & 1 223 614 & 13 & 2 578 & 16 562 575 \\ \cline{2-8} 
\multicolumn{1}{|c|}{} & \begin{math} \geqslant \end{math} 400 & 10 & 619 & 1 174 345 & 13 & 2 198 & 16 431 086 \\ \cline{2-8} 
\multicolumn{1}{|c|}{} & \begin{math} \geqslant \end{math} 500 & 10 & 518 & 1 129 434 & 13 & 1 956 & 16 322 797 \\ \hline
\multicolumn{1}{|c|}{\multirow{5}{*}{CC}} & \begin{math} \geqslant \end{math} 50 & 9 & 460 & 332 302 & 10 & 993 & 4 121 247 \\ \cline{2-8} 
\multicolumn{1}{|c|}{} & \begin{math} \geqslant \end{math} 100 & 7 & 329 & 323 046 & 10 & 732 & 4 102 610 \\ \cline{2-8} 
\multicolumn{1}{|c|}{} & \begin{math} \geqslant \end{math} 200 & 6 & 211 & 306 914 & 10 & 538 & 4 075 883 \\ \cline{2-8} 
\multicolumn{1}{|c|}{} & \begin{math} \geqslant \end{math} 300 & 6 & 160 & 294 747 & 9 & 433 & 4 050 072 \\ \cline{2-8} 
\multicolumn{1}{|c|}{} & \begin{math} \geqslant \end{math} 400 & 5 & 129 & 283 879 & 9 & 366 & 4 027 032 \\ \cline{2-8} 
\multicolumn{1}{|c|}{} & \begin{math} \geqslant \end{math} 500 & 5 & 93 & 268 132 & 8 & 327 & 4 009 762 \\ \hline
\end{tabular}
\end{table}
\pagebreak
\subsection{Evaluation of GO Level Specific performances}
Level specific performance results (y-axis) for different datasets (x-axis) are given as line plots and box plots in Figure 4 with individual plots for each GO category. As observed from Figure 4, F-score values are highly variable between 0.15 and 1.0 for different models. As observed from line plots (Figure 4A), there is no correlation between GO levels and performance of the system; however, increasing the training set sizes elevates the classification performance for all GO categories. Box plots (Figure 4B) also show that the performance variance at each GO level decreases with the increasing training set sizes, for MF and CC categories.
\begin{figure}[h]
  \centering
    \includegraphics[width=0.90\textwidth]{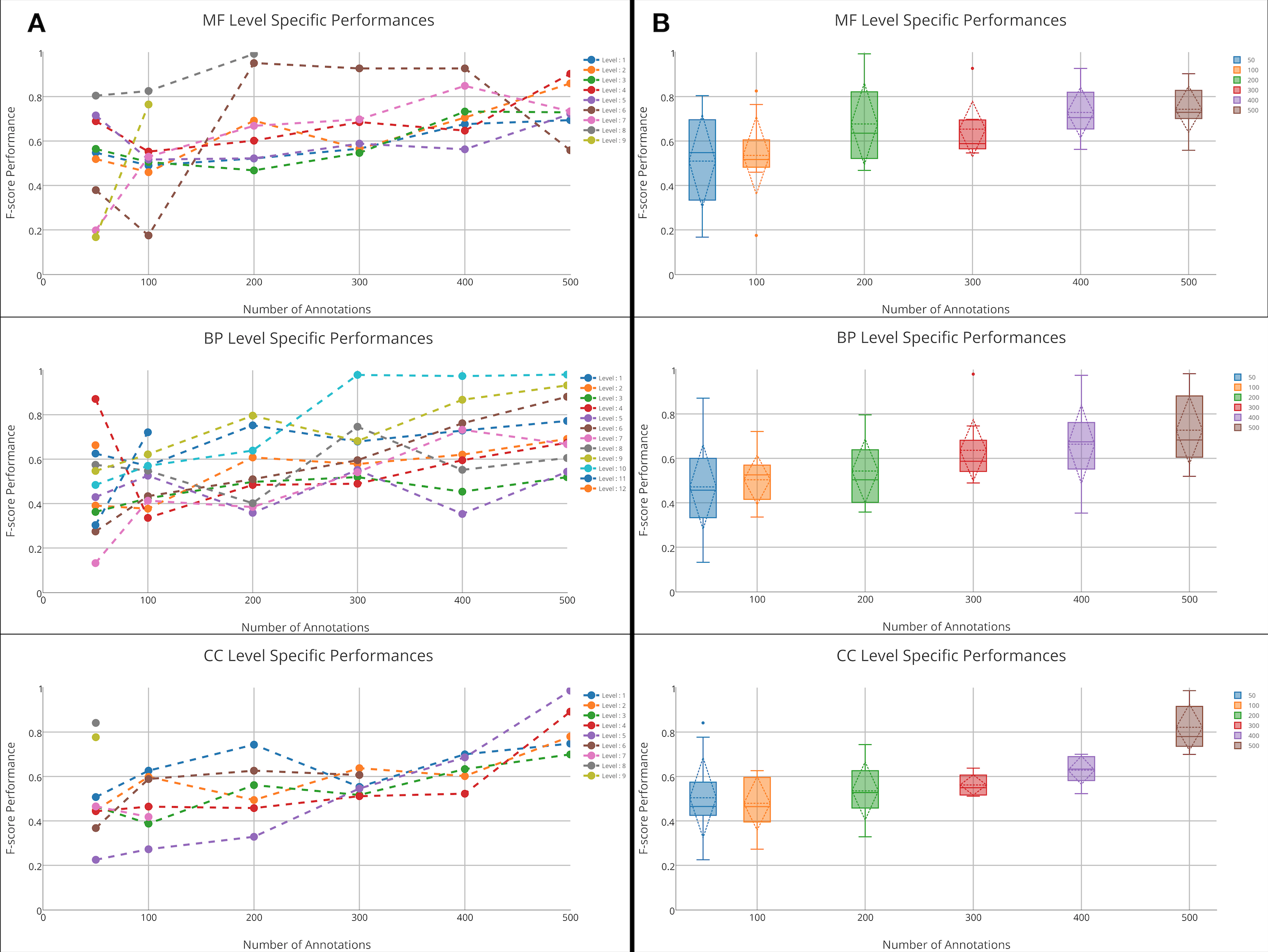}
      \caption{(A) Line plots displaying GO level specific performance for six training datasets for each GO category. (B) Box plots for level specific performance evaluation. Each box plot represents variance, mean and standard deviations of F-score values for different GO levels and training datasets, for each GO category. }
\end{figure}
In Figure 4, level specific performances were given and number of GO terms at each level highly varies among levels. Therefore, average GO term performances cannot be deducted from Figure 4. The average performances of models for each training set is given in Table 2. Each column in Table 2 corresponds to average F-Score values of GO terms belonging to a particular training dataset. The correlation between the training sample size and performance is also visible in this table.

\begin{table}[h]
\centering
\caption{The average performance (F-score) of GO terms belonging to different training datasets.}
\label{table-2}
\begin{tabular}{lllllll}
\hline
\multicolumn{1}{|l|}{} & \multicolumn{6}{l|}{\textbf{Training dataset sizes}} \\ \hline
\multicolumn{1}{|l|}{\textbf{GO categories}} & \multicolumn{1}{l|}{\textbf{\begin{math} \geqslant \end{math} 50}} & \multicolumn{1}{l|}{\textbf{\begin{math} \geqslant \end{math} 100}} & \multicolumn{1}{l|}{\textbf{\begin{math} \geqslant \end{math} 200}} & \multicolumn{1}{l|}{\textbf{\begin{math} \geqslant \end{math} 300}} & \multicolumn{1}{l|}{\textbf{\begin{math} \geqslant \end{math} 400}} & \multicolumn{1}{l|}{\textbf{\begin{math} \geqslant \end{math} 500}} \\ \hline
\multicolumn{1}{|l|}{\textbf{Molecular Function}} & \multicolumn{1}{r|}{0.69} & \multicolumn{1}{r|}{0.72} & \multicolumn{1}{r|}{0.71} & \multicolumn{1}{r|}{0.76} & \multicolumn{1}{r|}{0.82} & \multicolumn{1}{r|}{0.82} \\ \hline
\multicolumn{1}{|l|}{\textbf{Biological Process}} & \multicolumn{1}{r|}{0.42} & \multicolumn{1}{r|}{0.42} & \multicolumn{1}{r|}{0.47} & \multicolumn{1}{r|}{0.47} & \multicolumn{1}{r|}{0.50} & \multicolumn{1}{r|}{0.51} \\ \hline
\multicolumn{1}{|l|}{\textbf{Cellular Component}} & \multicolumn{1}{r|}{0.63} & \multicolumn{1}{r|}{0.65} & \multicolumn{1}{r|}{0.66} & \multicolumn{1}{r|}{0.69} & \multicolumn{1}{r|}{0.70} & \multicolumn{1}{r|}{0.73} \\ \hline
\end{tabular}
\end{table}
\subsection{Evaluation of the performance when the training sets are enriched with electronic annotations}
In UniProtKB/SwissProt, only 1\% of annotations have manual experimental evidence codes. The remaining annotations have the other 15 evidence codes and these annotations are considered as less reliable annotations. Here, we investigated the performance change when all annotations were included in training. Our first objective was to investigate whether deep learning algorithms can handle noisy data. Second, we wanted to observe how the performances of GO terms were affected when training was performed using annotations with all evidence codes. To perform this experiment, we first chose the MF GO terms whose annotation count was increased at least four times when annotations with all evidence codes were added. There were 20 MF GO terms which satisfied this condition. Subsequently, training was performed with the updated training data and individual performances of the GO terms were evaluated. Training data size and performance values are given in Table 3. We divided GO terms into two main categories as high performance GO terms and low performance GO terms based on their performances when the system was trained only with annotations having manual experimental evidence codes. Rows highlighted with bold characters represent the GO terms whose performances were significantly decreased and increased, respectively. The remaining GO terms are the ones for which we did not observe significant performance change. Results showed that performing training with all annotations significantly increased the performances of the low performed GO terms. However, including all annotations in training of high performed GO terms decreased their performances in some of the investigated cases. 

\begin{table}[h]
\centering
\caption{Performance (F-score) changes for selected GO terms after the enrichment of training sets.}
\label{table-3}
\begin{tabular}{l|l|r|r|r|r|r|}
\cline{2-7}
 & \multicolumn{1}{c|}{\textbf{GO Term}} & \multicolumn{1}{c|}{\textbf{\begin{tabular}[c]{@{}c@{}}NoA*\\ (MEE*)\end{tabular}}} & \multicolumn{1}{c|}{\textbf{\begin{tabular}[c]{@{}c@{}}NoA \\ (AE*)\end{tabular}}} & \multicolumn{1}{c|}{\textbf{\begin{tabular}[c]{@{}c@{}}F-score \\ (MEE)\end{tabular}}} & \multicolumn{1}{c|}{\textbf{\begin{tabular}[c]{@{}c@{}}F-score \\ (AE)\end{tabular}}} & \multicolumn{1}{c|}{\textbf{\begin{tabular}[c]{@{}c@{}}Performance \\ change\end{tabular}}} \\ \hline
\multicolumn{1}{|l|}{\multirow{7}{*}{\textbf{\begin{tabular}[c]{@{}l@{}}Initially low \\ performance \\ GO terms\end{tabular}}}} & \textbf{GO:0097367} & \textbf{1 413} & \textbf{7 377} & \textbf{0.32} & \textbf{0.79} & \textbf{0.47} \\ \cline{2-7} 
\multicolumn{1}{|l|}{} & \textbf{GO:0043167} & \textbf{2 313} & \textbf{11 549} & \textbf{0.36} & \textbf{0.72} & \textbf{0.36} \\ \cline{2-7} 
\multicolumn{1}{|l|}{} & \textbf{GO:0036094} & \textbf{1 723} & \textbf{8 345} & \textbf{0.44} & \textbf{0.78} & \textbf{0.34} \\ \cline{2-7} 
\multicolumn{1}{|l|}{} & \textbf{GO:0030554} & \textbf{675} & \textbf{5 540} & \textbf{0.59} & \textbf{0.92} & \textbf{0.33} \\ \cline{2-7} 
\multicolumn{1}{|l|}{} & \textbf{GO:0032555} & \textbf{932} & \textbf{6 481} & \textbf{0.61} & \textbf{0.91} & \textbf{0.30} \\ \cline{2-7} 
\multicolumn{1}{|l|}{} & \textbf{GO:0000166} & \textbf{1 440} & \textbf{7 759} & \textbf{0.56} & \textbf{0.85} & \textbf{0.29} \\ \cline{2-7} 
\multicolumn{1}{|l|}{} & \textbf{GO:0043169} & \textbf{1 918} & \textbf{10 885} & \textbf{0.50} & \textbf{0.74} & \textbf{0.24} \\ \hline
\multicolumn{1}{|l|}{\multirow{13}{*}{\textbf{\begin{tabular}[c]{@{}l@{}}Initially high \\ performance \\ GO terms\end{tabular}}}} & GO:1901265 & 1 440 & 7 760 & 0.78 & 0.81 & 0.03 \\ \cline{2-7} 
\multicolumn{1}{|l|}{} & GO:0005524 & 584 & 5 389 & 0.91 & 0.93 & 0.02 \\ \cline{2-7} 
\multicolumn{1}{|l|}{} & GO:0035639 & 820 & 6 325 & 0.91 & 0.91 & 0.00 \\ \cline{2-7} 
\multicolumn{1}{|l|}{} & GO:0032559 & 659 & 5 515 & 0.89 & 0.89 & 0.00 \\ \cline{2-7} 
\multicolumn{1}{|l|}{} & GO:0001883 & 871 & 6 364 & 0.93 & 0.90 & -0.03 \\ \cline{2-7} 
\multicolumn{1}{|l|}{} & GO:0032549 & 872 & 6 373 & 0.93 & 0.90 & -0.03 \\ \cline{2-7} 
\multicolumn{1}{|l|}{} & GO:0046872 & 1 868 & 10 783 & 0.87 & 0.79 & -0.07 \\ \cline{2-7} 
\multicolumn{1}{|l|}{} & \textbf{GO:0001882} & \textbf{879} & \textbf{6 397} & \textbf{0.95} & \textbf{0.84} & \textbf{-0.11} \\ \cline{2-7} 
\multicolumn{1}{|l|}{} & \textbf{GO:0008270} & \textbf{471} & \textbf{2 723} & \textbf{0.82} & \textbf{0.71} & \textbf{-0.11} \\ \cline{2-7} 
\multicolumn{1}{|l|}{} & \textbf{GO:0005525} & \textbf{254} & \textbf{1 047} & \textbf{0.96} & \textbf{0.82} & \textbf{-0.15} \\ \cline{2-7} 
\multicolumn{1}{|l|}{} & \textbf{GO:0032550} & \textbf{868} & \textbf{6 357} & \textbf{0.93} & \textbf{0.70} & \textbf{-0.23} \\ \cline{2-7} 
\multicolumn{1}{|l|}{} & \textbf{GO:0017076} & \textbf{956} & \textbf{6 510} & \textbf{0.93} & \textbf{0.68} & \textbf{-0.25} \\ \cline{2-7} 
\multicolumn{1}{|l|}{} & \textbf{GO:0032553} & \textbf{999} & \textbf{6 622} & \textbf{0.92} & \textbf{0.65} & \textbf{-0.27} \\ \hline
 & \textbf{Average} & 1 098 & 6 805 & 0.75 & 0.81 & 0.06 \\ \cline{2-7} 
\end{tabular}
\end{table}
\pagebreak
\subsection{Evaluation of Overall System Performance} 
In order to calculate the overall evaluation of the performance, we considered the system as a whole and applied the hierarchical evaluation method that was described in the previous section. We evaluated the overall system performance using the models that were trained by the GO terms which have at least 500 annotations. We used a separate test dataset that was not used in the training phase of the system. Test dataset includes both positive and negative protein sequences for each model. We fed all test proteins to all the models and evaluated the system performance using Matthews correlation coefficient. We first evaluated the performance without using hierarchical evaluation method and the performance was calculated as 0.46, 0.34 and 0.32 for molecular function, biological process and cellular component categories, respectively. When we employed the hierarchical evaluation method, the overall system performance was calculated as 0.75, 0.49 and 0.63 for molecular function, biological process and cellular component categories, respectively.
\begin{table}[]
\centering
\caption{The average performance results with and without the hierarchical evaluation procedure.}
\label{table-4}
\begin{tabular}{|c|r|r|r|r|r|r|}
\hline
\multicolumn{1}{|l|}{} & \multicolumn{3}{c|}{\textbf{\begin{tabular}[c]{@{}c@{}}Hierarchical \\ Evaluation\end{tabular}}} & \multicolumn{3}{c|}{\textbf{\begin{tabular}[c]{@{}c@{}}Without \\ Hierarchical Evaluation\end{tabular}}} \\ \hline
\multicolumn{1}{|l|}{} & \multicolumn{1}{c|}{\textbf{MCC}} & \multicolumn{1}{c|}{\textbf{Precision}} & \multicolumn{1}{c|}{\textbf{Recall}} & \multicolumn{1}{c|}{\textbf{MCC}} & \multicolumn{1}{c|}{\textbf{Precision}} & \multicolumn{1}{c|}{\textbf{Recall}} \\ \hline
\textbf{\begin{tabular}[c]{@{}c@{}}Molecular \\ Function\end{tabular}} & 0.75 & 0.81 & 0.94 & 0.46 & 0.68 & 0.85 \\ \hline
\textbf{\begin{tabular}[c]{@{}c@{}}Biological \\ Process\end{tabular}} & 0.49 & 0.72 & 0.80 & 0.34 & 0.57 & 0.95 \\ \hline
\textbf{\begin{tabular}[c]{@{}c@{}}Cellular \\ Component\end{tabular}} & 0.63 & 0.90 & 0.61 & 0.32 & 0.60 & 0.90 \\ \hline
\end{tabular}
\end{table}
\pagebreak
\section{Discussion}
\label{sec:alg}
In this study, we described DEEPred method for predicting protein functions using multi-task deep neural networks. We trained several DNN models using 6 training datasets containing GO terms with differing number of training samples and the best performing models were selected. Our level specific (Figure 4) and overall performance (Table 4) evaluation results showed that there is a general trend of performance increase with the increasing number of training samples which means that including GO terms with small number of protein associations in our models decreases the overall performance (Figure 4). Therefore, training data size is one of the key factors that affects the performance in deep learning algorithms. We also investigated if there is a relationship between levels of GO terms (i.e. GO terms describing specific functions at the leaf nodes as opposed to terms describing broad functions close to the root of the GO graph) on a GO DAG and their classification performances. Figure 4A (line plots) showed that there is no correlation between GO levels and classification performance. In addition, we observed that the variance of the performances between different GO levels decreases as the training dataset size increases for molecular function and cellular component categories (Figure 4B). For biological process category, performance also increases with increasing GO training dataset sizes, however the variance is relatively higher, whose main reason may be attributed to the biological process GO terms representing metabolic processes (e.g. GO:0006099 - tricarboxylic acid cycle) that involves several events which is hard to associate with a sequence signature. Figure 4B also showed that performance variances of cellular component GO terms is less than the molecular function and biological process category. The reason for observing less variance in cellular component category could be that the hierarchy between cellular comportment GO terms (cellular comportments) is inherently available within cells which results in better defined hierarchical relationships between cellular component GO terms.
Overall performance evaluation results indicated that the performance of the proposed architecture was satisfactory (Table 4). Since the system considers the prediction scores of the parents of a GO term for the query protein in order to avoid false positive hits (i.e. the hierarchical evaluation procedure), the provided predictions can be considered as reliable, and this is reflected on the performance values as elevated precision measurements (Table 4).
\\
In most of the protein function prediction methods, training is performed using only the annotations with experimental evidence codes in order to avoid potential error propagation. The disadvantage of this approach is that most GO terms have a few number of annotated proteins, which is not sufficient for a machine learning model training. Therefore, the functions defined by these terms cannot be predicted by automated methods. One solution would be including the annotations with non-experimental evidence codes such as the electronic annotations (i.e. the annotations produced by other automated methods). For example, the number of GO terms that have more than 50 protein associations is calculated as 627 when we only considered the annotations with manual experimental evidence codes. However, when we considered the annotations with all evidence codes, the number of GO terms that have more than 50 protein associations became 2143. Thus, if the annotations with all evidences are included, we can provide predictions for more GO terms. The downside of adding annotations with non-experimental evidence codes to the training dataset is that false positive samples may have been incorporated into the training sets, which would result in error propagation. We investigated the performance change when the annotations with all evidence codes were included for training, and its performance results were compared to the ones when the training was performed using only the annotations with manual experimental evidence codes (Table 3). Evaluation results showed that the accuracy over the initially low performed GO terms were increased significantly, which indicates that deep learning algorithms are tolerant to noise in the learning data. Therefore, annotations from other evidence codes can be included in the training of low performed GO terms, where there is still room for significant performance improvement. However, including less reliable annotations in the training set of initially high performed GO terms decreased the performance for a portion of them. Nevertheless, there was no room for any performance improvement for these GO terms.
\\
Finally, we employed a hierarchical evaluation method to provide more accurate predictions by taking the prediction scores of the parents of the target GO term into account, along with the score of the target term. The evaluation results indicated that the recall values were relatively decreased and the precision scores were increased when we employed the hierarchical evaluation procedure, resulting in an increased overall performance (i.e. MCC) of the system (Table 4). The reason for such performance results was that the number of false positive hits decreased significantly, which was a significant factor affecting the overall performance before the employment of hierarchical evaluation. 

\section{Conclusions}
\label{sec:concl}
Deep learning algorithms have been shown to be useful in various fields and shown to enhance the prediction results, however, to the best of our knowledge it was not applied to protein function prediction area at a large-scale. Here, we proposed a hierarchical deep neural network architecture for the prediction of functions using GO terms. The proposed study had three main objectives: \textit{(i)} to investigate the potential of deep learning techniques for protein function prediction; \textit{(ii)} to reveal the relationship among the performance of the system and various parameters of the system such as the size of the training datasets for each class, functional term specificity in terms of the different levels of GO; and \textit{(iii)} to investigate the performance change when the annotations with low reliability (i.e. the ones that possess non-experimental evidence codes) are also included in the training of the system, in order to enrich the training sets.
\\
The proposed level specific architecture is successful to discriminate the functions of proteins. In addition, considering the parents of the GO terms when producing the predictions yielded a higher performance in terms of precision. Our results demonstrated that deep learning can be employed to significantly improve the performance of prediction (F-score \begin{math} \geqslant \end{math} 0.75) for hard to predict GO categories such as the biological process and the cellular component, especially when the training set sizes are sufficiently large (> 500 proteins). Average F-Score values showed that including GO terms with lower number of protein associations decreases the overall performance. Since deep learning algorithms are reported to be able to handle noisy data, less reliable annotations can be incorporated to the training to elevate the accuracy of the GO terms with low performances; however, this addition decreases the performance of high-performing models in some cases. It was also displayed that the overall system performance increases by performing the proposed hierarchical evaluation procedure. As a conclusion, we showed that deep learning techniques has a significant potential in automated protein function prediction. 
\\
As future work, our aim is to further investigate the model behavior under different circumstances and to optimize the models to provide DEEPred as an open-access tool to the research community. We also plan to apply deep learning algorithms for automated protein function prediction using feature vectors obtained from protein descriptors as well as from various types of functional annotations from other biological data sources that are cross-referenced in UniProt database (i.e. structural information, PPIs, phylogeny, protein families/domains and classifications, gene expression data, chemistry information, etc.). The features playing a critical role in the predictions will be further investigated and will be selected in combination to finalize the models. The trained models will be employed to produce large scale functional associations for the whole recorded protein space using GO terms, EC numbers, keywords, comments, pathway information and others.

\section{References}
\label{sec:ref}
\begin{enumerate} 
    \item UniProt Consortium. UniProt: the universal protein knowledgebase. Nucleic Acids Res. 2016;45:1–12.
\item Blake JA, Christie KR, Dolan ME, Drabkin HJ, Hill DP, Ni L, et al. Gene ontology consortium: Going forward. Nucleic Acids Res. 2015;43:D1049–56. 
\item Doğan T, MacDougall A, Saidi R, Poggioli D, Bateman A, O’Donovan C, et al. UniProt-DAAC: domain architecture alignment and classification, a new method for automatic functional annotation in UniProtKB. Bioinformatics. 2016;32:2264–71.
\item Lan L, Djuric N, Guo Y, Vucetic S. MS-kNN: protein function prediction by integrating multiple data sources. BMC Bioinformatics. 2013;14:1–10.
\item Wass MN, Barton G, Sternberg MJE. CombFunc: predicting protein function using heterogeneous data sources. Nucleic Acids Res. 2012;40:W466-70. 
\item Tiwari AK, Srivastava R. A survey of computational intelligence techniques in protein function prediction. Int. J. Proteomics. 2014;1:1-22.
\item Koskinen P, Törönen P, Nokso-Koivisto J, Holm L. PANNZER: High-throughput functional annotation of uncharacterized proteins in an error-prone environment. Bioinformatics. 2015;31:1544–52. 
\item Jiang Y, Oron TR, Clark WT, Bankapur AR, D’Andrea D, Lepore R, et al. An expanded evaluation of protein function prediction methods shows an improvement in accuracy. Genome Biol. 2016;17:1–19.
\item Radivojac P, Clark WT, Oron TR, Schnoes AM, Wittkop T, Sokolov A, et al. A large-scale evaluation of computational protein function prediction. Nat. Methods. 2013;10:221–9. 
\item Hinton G, Deng L, Yu D, Dahl GE, Mohamed A, Jaitly N, et al. Deep Neural Networks for Acoustic Modeling in Speech Recognition. IEEE Signal Process. Mag. 2012;82–97. 
\item Deng L, Hinton G, Kingsbury B. New Types of Deep Neural Network Learning For Speech Recognition And Related Applications.: An Overview. 2013;1–5.
\item Angermueller C, Pärnamaa T, Parts L, Oliver S. Deep Learning for Computational Biology. Mol. Syst. Biol. 2016;12:1–16. 
\item Min S, Lee B, Yoon S. Deep Learning in Bioinformatics. Brief Bioinform. 2016;bbw068:1–46. 
\item Taigman Y, Ranzato MA, Aviv T, Park M. Deepface : Closing the Gap to Human-Level Performance in Face Verification. 2014;1–8. 
\item Lecun Y, Bengio Y, Hinton G. Deep learning. Nature. 2015;521:436–44. 
\item Gawehn E, Hiss JA, Schneider G. Deep Learning in Drug Discovery. Mol. Inform. 2016;35:3–14. 
\item Baskin II, Winkler D, Tetko I V. A renaissance of neural networks in drug discovery. Expert Opin. Drug Discov. 2016;11:785–95. 
\item Mayr A, Klambauer G, Unterthiner T, Hochreiter S. DeepTox.: Toxicity Prediction using Deep Learning. Front. Environ. Sci. 2016;3:1–15. 
\item Ramsundar B, Riley P, Webster D, Konerding D, Edu KS, Edu PS. Massively Multitask Networks for Drug Discovery arXiv:1502.02072v1. arXiv. 2015;1–27. 
\item Bengio Y. Learning Deep Architectures for AI. Found. Trends Mach. Learn. 2009; 2:1-127. 
\item Pérez-sianes J, Pérez-sánchez H, Díaz F. Virtual Screening.: A Challenge for Deep Learning. 10th Int. Conf. PACBB, Adv. Intell. Syst. Comput. 2016;13–22. 
\item Sliwoski G, Kothiwale S, Meiler J, Lowe EW. Computational Methods in Drug Discovery. Pharmacol. Rev. 2014;66:334–95. 
\item Kotsiantis SB, Kanellopoulos D, Pintelas PE. Data preprocessing for supervised learning. Int. J. Comput. Sci. 2006;1:111–7.
\item Bruha I, Famili A. Postprocessing in machine learning and data mining. ACM SIGKDD Explor. Newsl. 2000. 
\item Sarac OS, Gürsoy-Yüzügüllü O, Cetin-Atalay R, Atalay V. Subsequence-based feature map for protein function classification. Comput. Biol. Chem. 2008.;32:122–30.
\item Hinton GE, Osindero S, Teh YW. A fast learning algorithm for deep belief nets. Neural Comput. 2006;18:1527–54.
\item Srivastava N, Hinton G, Krizhevsky A, Sutskever I, Salakhutdinov R. Dropout: A Simple Way to Prevent Neural Networks from Overfitting. J. Mach. Learn. Res. 2014;15:1929–58. 
\item Abadi M, Agarwal A, Barham P, Brevdo E, Chen Z, Citro C, et al. TensorFlow: Large-Scale Machine Learning on Heterogeneous Distributed Systems. ArXiv. 2015;1:1–19.

\end{enumerate}

\bibliographystyle{plain}
\bibliography{references}       

\end{document}